\documentclass[12pt]{sadhana}

\usepackage{graphicx,amsmath,bm}

\usepackage[numbers,super]{natbib}
\usepackage{setspace}
\makeatletter
\renewcommand\@makefnmark{}
\makeatother

\onecolumn
\begin{document}

\title{GPS in the cosmos}


\author{Gopal-Krishna\textsuperscript{1,*}}
\affilOne{\textsuperscript{1} UM-DAE Centre for Excellence in Basic Sciences, Vidyanagari, Mumbai 400098, India}


\maketitle

\newcommand{\uniformfontsize}{\fontsize{12pt}{14pt}\selectfont}

\makeatletter
\renewcommand\section{\@startsection {section}{1}{\z@}%
                                   {-3.5ex \@plus -1ex \@minus -.2ex}%
                                   {2.3ex \@plus.2ex}%
                                   {\normalfont\uniformfontsize\bfseries}}

\makeatother

\begin{abstract}
This note chronicles the early steps and incidences from over four decades ago, that gave a kickstart to research on the now prominent subclass of radio galaxies, called `Gigahertz-Peaked-Spectrum' (GPS) sources. In this first-hand account, the origin of the acronym `GPS' in this context is retraced, along with the first crucial steps and the major milestones set in course of evolution of this now vibrant branch of extragalactic radio astronomy. A brief update on the recent developments in this field is also presented.
\end{abstract}

\msinfo{XX July 2024}{XX YY 2024}{XX YY 2024}

\keywords{Galactic and extra-galactic astrophysics — Radio continuum: galaxies — Galaxies: active — Galaxies: jets — History of astronomy\\ 
``What’s in a name? That which we call a rose/ By any other name would smell as sweet." — Shakespeare}

\articleType{}
\vspace{1cm}
\section{Background}
\label{introduction}
Systematic measurements of radio-frequency spectra of a large number of (strong) extragalactic radio sources began \cite{1969ApJ...157....1K} in the 1960s, 
mainly targeting the 3rd Cambridge (3C) catalogue\cite{1962MmRAS..68..163B} which was essentially complete above a flux density limit of 9 Jy at 178 MHz. The measured approximately straight radio spectra of most of those sources, with a spectral index ($\alpha$) of around $-$ 0.75, provided a crucial independent evidence supporting the synchrotron mechanism generating their radio emission, as originally proposed by Pikel'ner\cite{Pikelner1953DoSSR..88..229P} and Shklovskii\cite{Shklovskii1953AZh....30...15S,Shklovskii1955AZh....32..215S}. But, even in those early studies, a small fraction of extragalactic sources was found to exhibit complex radio spectra marked by one or more peaks, or simply a flattish profile, often turning over at metre wavelengths. A rare, striking deviant from the canonical straight spectrum was the southern radio source 1934-63; its spectrum showed a single, sharp peak near 1.5 GHz and sloping downward on either side\cite{Kellermann1966AuJPh..19..195K,1963Natur.199..682B,1965Natur.206..176S}. As described below, that spectrum was to become the prototype of the subclass of radio galaxies/quasars which have, since the mid-1980s, come to be known as `Gigahertz-Peaked-Specrtum' (GPS) sources and have acquired a prominent place among the zoo of extragalactic radio sources (see the article by Kellermann\cite{2012arXiv1210.0984K}). Over 8000 GPS sources have now been catalogued and they are estimated to be $\sim 3\%$ of extragalactic radio source population at 1.4 GHz\cite{2024arXiv:2406.13346}. 
This article attempts to chronicle the events that not only led to the GPS nomenclature but, more importantly, shifted this research topic into a high gear around the mid-1980s. Subsequently, a few variants of the GPS sources have come to be recognised. Prominent among these are `High-Frequency-Peakers' (HFPs\cite{2000A&A...363..887D}) and `Megahertz-Peaked-Spectrum' sources (MPS\cite{2015MNRAS.450.1477C}). Recently, all these subclasses, including GPS, have been subsumed within a common rubric, namely `Peaked-Spectrum-Sources' (`PS', see the review by O'Dea \& Saikia\cite{2021A&ARv..29....3O}; also, Annexure II).\footnote{{$^*$E-mail: gopaltani@gmail.com}}

As discussed in Kellermann\cite{Kellermann1966AuJPh..19..195K}, three principal mechanisms considered for the sharp spectral turnover of the source 1934-63 below $\sim$ 1 GHz are: synchrotron self-absorption (SSA), a dispersive medium\cite{Razin1960} and free-free absorption (FFA). On a lighter note, the sharply peaked radio spectrum of 1934-63 appeared so odd, at the first sight, that a reference was made in that paper to its similarity to the spectral shape predicted by Kardashev\cite{1964SvA.....8..217K} for signals transmitted by extraterrestrial `super-civilisations'. However, it did not take long for this speculation to be put to rest, with the identification of the source with a galaxy showing an emission-line rich (quasar-like) optical spectrum corresponding to a redshift $z$ = 0.182\cite{1972ApJ...173L..19P}.

Curiously, this exciting debut was followed by a relative lull in the discovery of more sources showing a peaked radio spectrum. A breakthrough came a decade later, with the advent of high-fidelity Very-Long-Baseline-Interferometry (VLBI) imaging, through the development of a `hybrid' imaging technique which combined the closure-phase information with fringe amplitudes\cite{1974ApJ...193..293R,1977Natur.269..764W}. An unexpected outcome from this was that the radio source CTD93, instead of being a compact source, as expected from its single-peaked radio spectrum (e.g., \cite{Kellermann1966AuJPh..19..195K,1965Natur.206..176S}), was clearly resolved into a roughly symmetric pair of tiny radio components separated by approximately 10$^2$ parsec\cite{1980ApJ...236...89P,1982A&A...106...21P}. Thus, its morphology appeared to be a miniature replica of the $\sim$ 10$^3$ time larger classical double radio source Cygnus A, the tantalizing discovery made by Jennison \& Das Gupta\cite{1953Natur.172..996J}. Quite aptly, CTD93 was dubbed `compact double' (CD) and deemed a baby version and a likely progenitor of classical double radio sources (\cite[see,][]{1980ApJ...236...89P,1981ApJ...244...19P,1981AJ.....86.1600M}). By 1982, the list of such (symmetric) compact double sources showing a clear centimetre-excess spectrum peaking at about 1 GHz had inched to nearly half-a-dozen\cite{1982A&A...106...21P,1969ApJ...155L..71K}, albeit lagging far behind the burgeoning interest among the large community of radio galaxy researchers in such `nascent' Cygnus A radio sources. Some additional finer-tuned classifications were subsequently proposed, such as `compact triples'\citealp{1992ApJ...396...62C}, and `compact-symmetric-objects' (CSOs, see \citealp{1994ApJ...432L..87W,2001A&A...369..380F}). Recently, CSOs have been argued to be mostly `short-lived' events leading to radio lobes that grow no bigger than $\sim$ 1 kpc\citealp{Kiehlmann2024ApJ...961..241K,1994cers.conf...17R}, rather than being `young' progenitors of standard double radio sources, as was inferred in many earlier works, e.g., \citealp{1995A&A...302..317F,1996ApJ...460..634R,2005ApJ...622..136G}, see also, \citealp{1997ApJ...487L.135R,2006A&A...450..945K}.

The tardy progress until early 1980s, as mentioned above, gave this author the impetus to undertake a systematic search for CD candidates, mainly diagnosed by a single-peaked radio spectrum attaining a maximum around 1 GHz. A second motivating factor was the result newly highlighted by Peterson et al.\cite{1982ApJ...260L..27P} who showed that radio spectra of all the distant ($z > 3$) quasars then known, were dominated by a single peak seen below $\sim$ 8 GHz. Their list included the $z = 3.78$ quasar PKS 2000-33, holder of  the contemporary redshift record. It was evident (see, e.g., \citealp{1981MNRAS.197..593D,1981A&AS...43..381K}, also, \citealp{1978MNRAS.185..123K}) that such sources, exhibiting a spectral turnover at gigahertz frequencies would have been grossly under-represented in the existing major surveys most of which had been made at metre wavelengths and with relatively high flux density limits, which could well be the reason behind the afore-mentioned sluggish pace of discovery of CDs.

\section{End of the statis}

Motivated by the factors summarized above, our entry into this relatively virgin field in early 1980s started with a straight-forward, albeit rather ambitious goal of making a catalog of 50 sources whose spectrum is peaked around 1 GHz. Such a sample would mark an order-of-magnitude jump in the number of such sources known at that time and thus
be large enough to trigger and support a systematic research program to investigate this phenomenon. The endeavour began in 1981, on my return from an extended sabbatical visit to the Max-Planck-Institut f$\ddot{\mathrm{u}}$r Radioastronomie (MPIfR, Bonn), to my home institution, the Tata Institute of Fundamental Research (TIFR) in India, at its newly opened branch in Bangalore (now Bengaluru), called `TIFR Centre'. The project was initiated mainly with Dr Hans Steppe, my collaborator on the `BOOTY' (Bonn-Ooty) project (Annexure I) under which we had measured radio spectra of 577 extragalactic radio sources constituting an unbiased sample defined at the flux density level equivalent to that of the well-known large-area 4C survey which is believed to be essentially complete above a flux density of 2 Jansky at 178 MHz. A key advantage offered by the BOOTY sample was that the positions and sizes/structures of all its 577 sources had already been determined to arc-second precision, in course of our lunar occultation survey at 327 MHz, using India's large steerable cylindrical radio telescope at Ooty\cite{1971NPhS..230..185S,1971ApL.....9...53S}. In addition to the BOOTY sample, another contributor to our catalog of peaked-spectrum sources was the just released, large compendium of radio spectra of extragalactic sources, published in 1981 by an intrepid team consisting of Arno Witzel, Helmut K$\ddot{\mathrm{u}}$hr, Ivan Pauliny-Toth, and U. Nauber, all at MPIfR (Bonn).
Their catalog appeared in two releases: (i) a 1-Jansky catalog defined at 5 GHz\cite{1981A&AS...45..367K} and (ii) a less complete, but much larger compendium of extragalactic radio sources\cite{Kuehr1981MPIfR}.

{\bf Our first list:} Mainly using the above-mentioned databases together with other sufficiently precise flux-density measurements scattered through the literature, we were able to put together our first list of 25 sources with a spectrum marked by a single peak within the frequency range 0.5 to 5 GHz. In this undertaking, Dr Steppe and I were joined by Mr Alok Patnaik, then a graduate student at the TIFR Centre (Bangalore). Our paper titled {\it `A sample of 25 extragalactic radio sources having a spectrum peaked around 1 GHz'}, following a prompt approval by the reviewer Philippe V{\'e}ron, was published in {\it Astronomy \& Astrophysics (1983)}, authored by Gopal-Krishna, Patnaik, A. \& Steppe, H\cite{1983A&A...123..107G}.

Although we had coined the term `Gigahertz-Peaked-Spectrum' (GPS) for the sources in our first list itself, we desisted from using it in that paper, following the realisation that the 3 letters in GPS were also the first letters of the names of the 3 authors of that paper (List 1), as mentioned in the preceding sentence. We therefore felt that calling these sources `GPS' in our first list itself would tantamount to eponymy and appear immodest. Nonetheless, in 1983 itself, we had begun using the notations `gigahertz-peaked-spectrum' sources, and `peaked-spectrum' sources, as seen from the 1983 notifications we received from the Effelsberg Radio Telescope `Time Allocation Committee', signed by Prof. Ivan Pauliny-Toth and another one signed by Dr R. Schwartz (Annexure II).

{\bf The second and third lists:} In order to realise our original goal of publishing a sample of 50 GPS sources, a second list\cite{1985A&A...152...38S} 
containing another 25 radio sources with a single-peaked-spectrum (near 1 GHz) was published. In that list, we formally introduced the term {\bf Gigahertz-Peaked-Spectrum (GPS)}. As for our first list, the candidates for this second list were also drawn from the literature. However, in order to confirm the spectral peaking of 6 of the shortlisted candidates, their flux densities needed to be measured at 1.4/0.3 GHz. To do this, I invited participation of Dr T.A. Th. Spoelstra (Netherland Foundation for Radio Astronomy) with whom I had become acquainted during his visit to our observatory at Ooty during 1982. Titus rendered valuable help in the preparation of the GPS list 2 (and even our GPS list 3), by observing several candidate GPS sources with the Westerbork Synthesis Radio Telescope (NOTE: Our list-3 containing another 10 GPS sources was 
published\cite{1993A&A...271..101G} almost a decade after the first two lists mentioned above).

\section{The kickstart and thereafter} 

In addition to being `mini-Cygnus A' candidates, or high-$z$ radio sources (Sect. \ref{introduction}), there were already hints that GPS sources might display only a mild flux variability and weak radio polarisation, properties then deemed uncharacteristic of the sources not exhibiting the canonical straight radio spectrum (e.g., \cite{1982A&A...106...21P,1982ApJ...255...39R,1984IAUS..110...15P}). As mentioned above, our first two lists containing a total of 50 GPS sources amounted to roughly an order-of-magnitude jump in the number of such sources known hitherto, thus boosting prospects of launching a systematic probe into the reasons behind their hinted abnormalities. Such a possibility cropped up, coincidentally, when in mid-1983, our co-author on both existing GPS lists, Alok Patnaik went to the National Radio Astronomy Observatory (NRAO, USA) to make VLA observations of objects related to his ongoing PhD work (in those early days of VLA, users were required to come to the VLA site to make their observations - then a tall order and indeed a major obstacle for interested VLA users based in India). So I suggested to Alok to utilise his VLA trip to explore the possibility of setting up a collaboration for the purpose of following-up our sample of peaked-spectrum sources in the just published List 1\cite{1983A&A...123..107G}, and also another 25 such sources in our List 2 which was nearing completion (see above). In his handwritten (pre-email era) letter dated 22-July-1983 from NRAO (USA), while updating on the progress of his work there, Alok wrote to me
(Annexure III):

``Our paper [GPS list 1] has appeared in June in A\&A. There is somebody here who would like to collaborate [with] us in VLA observations of these sources. He is Chris P. O'Dea, would be submitting his thesis on Head-Tail sources in September \& then join NRAO as a post doc. He says he would like to observe these at various VLA configurations and frequencies. I told him that it is interesting to study them in detail and will let him know after hear from you about this. He is very keen on the subject. Please let me know your reaction….''

{\it This letter is a classic example of the critical role chance plays in shaping/accelerating the evolution of a research area.}

In my quick response to Alok, I supported the idea enthusiastically and we provided to Chris O'Dea the various details of the 25 GPS sources in our just published List 1\cite{1983A&A...123..107G} and also provided him our yet-to-be-published List 2 which contained another 25 GPS sources\cite{1985A&A...152...38S}. Our 3rd list containing another 10 GPS sources\cite{1993A&A...271..101G} was also made available to Chris, well in advance of its publication. Subsequent events have proved this interaction to be a watershed, which gave a kickstart to the study of GPS sources, with Chris O'Dea emerging as the central player, particularly after the publication of his landmark review article in 1998, titled  {\it ``The Compact Steep-Spectrum and Gigahertz Peaked-Spectrum Radio Sources''}\cite{1998PASP..110..493O}.

In the interim, several variants of GPS sources have been conceived/recognised. These are chiefly distinguished by the radio-frequency segment within which the maximum of the single-peaked spectrum is found to occur, and/or on the basis of radio morphological properties\cite{2000A&A...363..887D,1994cers.conf...17R,1995A&A...302..317F,1998A&AS..131..303S,2000MNRAS.319..445S,2016MNRAS.459.2455C}. Consequently, the term `GPS' has been absorbed into a more general term `Peaked-Spectrum' (PS) sources (e.g., the recent review by O’Dea \& Saikia\cite{2021A&ARv..29....3O}; also, Annexure II), even though the nomenclature `GPS' continues to be in vogue. According to one viewpoint, these variants represent the same basic phenomenon at different evolutionary stages\cite{1998A&A...329..845C, Kunert2010MNRAS.408.2279K, 2012ApJ...760...77A}. A uniquely powerful handle became available from VLBI monitoring, leading to direct measurements of proper motion of the terminal hot spots in such 
mini-doubles\cite{ 1998A&A...336L..37O,1998A&A...337...69O,2021AN....342.1151O,2005ApJ...622..136G}. 
This has yielded the most reliable age estimates (kinematical ages, which are found to be as low as $\sim$ 100 years, or even lesser in some cases), making it possible to spot among the GPS sources the ones whose radio emission is not dominated by Doppler-boosted relativistic jets and which therefore look like genuine twin-lobed Cygnus A, in miniature. As mentioned above, this particular subset belongs to the category of `compact symmetric objects' (CSOs). Recently, it has been argued that most of them are too short-lived to evolve into classical double radio sources on the scale of 100 kpc (e.g., \cite{Kiehlmann2024ApJ...961..241K}). In some cases, however, it is conceivable that the advance of the hot spots gets quasi-stalled, conceivably due to the nascent jets ramming into outer parts of the torus, which have got tilted/distorted during gravitational interaction with an infalling massive body, like a dwarf galaxy\cite{1995PNAS...9211399G} (also, \cite{2009ApJ...697..1621G}). 
Some alternative physical processes behind this `jet-frustration' scenario have also been considered in the literature (e.g., \cite{1994ApJ...432L..87W,1982ApJ...262..529H,1984AJ.....89....5V,2022ApJ...927L..24G}). Simultaneous to blocking the advance of the radio jets, such an interaction with a (dense) interstellar medium (ISM) is also expected to boost the radio brightness
(via a better confinement of the radio-emitting plasmons), as well as lead to star formation in the main body of the host galaxy. Both possibilities were explored in Gopal-Krishna \& Wiita{\cite{1991ApJ...373..325G}}, where the first evidence for the star formation hypothesis was presented. This came from the observed bluer optical colour of the `compact-steep-spectrum' (CSS\cite{1981A&AS...43..381K,1982MNRAS.198..843P}) sources in the 3C catalog, in comparison to the optical colours of the hosts of much larger radio galaxies in the same catalogue.
Extensive VLBI monitoring campaigns would play a critical role in directly tracing out the kinematical evolution of individual CSOs and their progression towards the classical double radio sources. Important clues are also likely to emerge from high-resolution spectroscopic imaging, 
tracing the interaction of the nascent synchrotron jets with the ISM of the host galaxy (e.g., \cite{2023Galax..11...24M,Kukreti2024arXiv240706265K}).

Physical mechanisms behind the observed spectral turnover in the peaked-spectrum sources are still debated. A major clue in support of the synchrotron self-absorption (SSA) explanation comes from the strong anti-correlation observed for the peaked-spectrum radio sources, between linear size and the rest-frame spectral turnover frequency (\cite{2000MNRAS.319..445S,1990A&A...231..333F,1997AJ....113..148O,2009AN....330..214D,Jeyakumar2016MNRAS.458.3786J}, also, \cite{1996ApJ...460..612R,Tingay2003AJ....126..723T}). However, free-free absorption (FFA) remains a much-discussed competing alternative
(e.g., \cite{Kellermann1966AuJPh..19..195K,1997ApJ...485..112B,1999ApJ...521..103P,2003ApJ...584..135L,2018MNRAS.475.3493B,Sobolewska2019ApJ...871...71S,2019MNRAS.489.3506M,Shao2022A&A...659A.159S}, also, \cite{2014ApJ...780..178M,2015ApJ...809..168C,2019A&A...628A..56K}, see, for some early studies, \cite{1979ApJ...228...27M,1980PASA....4...70M,1981ApJ...247..419U,1984ApJ...277...82V}). A compelling evidence that FFA is at work would come from discovery of peaked-spectrum sources whose inverted radio spectrum after the spectral turnover towards lower frequencies becomes ultra-steep, with a spectral index breaching the theoretical upper limit of + 2.5 established for standard synchrotron sources. Note that although attainable only for an idealised, perfectly homogeneous radio source, this spectral index limit is practically independent of the slope of power-law energy distribution of the radiating relativistic electrons (e.g. \cite{1963Natur.199..682S,1979rpa..book.....R}, see also, \cite{1963Natur.200...56W}). Search for such sources, termed EISERS ({\it Extremely Inverted Spectrum Extragalactic Radio Sources}\cite{2014MNRAS.443.2824G}) has thrown up a handful of examples\cite{    2019MNRAS.485.2447M,2019MNRAS.489.3506M,2019A&A...628A..56K,2024arXiv:2406.13346}, but many more probably remain to be discovered, taking advantage of the recent sensitive decametric surveys, such as the LOFAR LBA Sky Survey at 54 MHz\cite{2021A&A...648A.104D}.






\appendix

\section*{Annexure I : The Bonn-Ooty (`BOOTY') sample of extragalactic radio sources}

The BOOTY sample of 577 extragalactic radio sources was assembled by me out of the lists of Ooty lunar occultation survey at 327 MHz, for the purpose of accurate flux density measurements at 2.7, 4.8 and, possibly 10.7 GHz, using the 100-meter Effelsberg radio telescope. These measurements were carried out mainly during 1979-80, at the Max-Planck-Institut f$\ddot{\mathrm{u}}$r Radioastronomie (Bonn), in collaboration with Dr Hans Steppe at the same institute. Going by the observing technique used, both the Ooty lunar occultation and the Effelsberg single-dish measurements enjoyed the advantage of having full sensitivity to radio structures on all relevant angular scales (i.e., freedom from the `missing' flux problem which often plagues interferometric observations). The source material for the BOOTY sample came from the Ooty lunar occultation (OTL) lists, starting from List 1\cite{1971ApL.....9...53S} to List 10\cite{1987A&AS...69...91S}, covering the observing period 1970-76. Median flux density of the OTL sample ($\sim$ 1 Jy at 327 MHz) is quite similar to that of the then available `deep' large-area survey of the northern sky, the 4C survey which is essentially complete above a 2-Jy limit at 178 MHz\cite{1965MmRAS..69..183P,1967MmRAS..71...49G}. While not a flux-limited complete sample, the source inclusion into the Ooty occultation lists can still be taken as unbiased, since it is based on the natural phenomenon of lunar occultation. One key advantage offered by this sample was its high positional accuracy (typically, of order of 1 arcsecond), which is almost two orders-of-magnitude superior to the 4C positions. Consequently, accurate flux-density measurements of the Ooty occultation sources at centimetre wavelengths (at 2.7 GHz for all 577 BOOTY sources and up to 10.7 GHz for 189 of them) could be accomplished quickly by taking `cross-scans' with the few-arcminute pencil-beam of the 100-metre Effelsberg dish. 

Thus, the Effelsberg single-dish flux measurements of the 
BOOTY sample defined at 327 MHz, yielded decimetric-wave spectral indices of a large number of relatively weak (i.e., 4C level) radio sources selected at metre wavelengths, with a precision comparable to what had hitherto been achieved routinely only for the radio spectra of an order-of-magnitude stronger 3C radio sources 
(e.g., \cite{1969ApJ...157....1K}). Furthermore, the availability of positional and angular-size/structural information with arc-second precision for the Ooty lunar occultation sources had already enabled a successful campaign to find their optical counterparts using the Palomar-Sky-Survey plates/prints, thanks mainly to the endeavours by Profs. M.N. Joshi and C.R. Subrahmanya (see, e.g., \cite{1971ApL.....9...53S,1987A&AS...69...91S,1973AJ.....78.1023J,1973AJ.....78..673K,1974AJ.....79..515K,1979MmASI...1....2S,1980MmASI...1...49J}). It may also be added that, with its selection at metre wavelength (327 MHz), the BOOTY sample is expected to be essentially free from orientation bias. The availability of accurate information on radio angular size, spectral index and optical identification, for the first time, for a large number of 4C-level radio sources, had delivered a number of interesting results, even before we set out to leverage our BOOTY sample for preparing the lists of GPS sources.  A few of those precursor findings are summarized below. \\

{\bf (a) `Very-steep-spectrum' radio sources (VSS, having $\alpha$ $<$ $-$ 1.1 at decimetre wavelengths)}

In a pioneering study, based on a sample of 4C radio sources imaged at 1.4 GHz with the 20'' beam of the Westerbork Synthesis Radio Telescope (WSRT), Tielens et al.\cite{1979A&AS...35..153T} compared the statistics of the optically identified fraction {\t vis a vis} radio angular size ($\theta$) for 17 VSS sources and 67 normal-spectrum radio sources. From the dependence of both these parameters on radio spectral index, these authors inferred that VSS sources are likely to be (statistically) more distant. Subsequently, this clue proved to be a stepping stone to highly effective future campaigns for discovering radio galaxies at very high redshifts (i.e., $z > 2$, going by the standards of those days) (see the review by Miley \& De Breuck\cite{2008A&ARv..15...67M}). The BOOTY sample provided the first independent check\cite{1981A&A...101..315G} to the Tielens et al. claim, using a sample which was not only 3 - 4 times larger than the 4C sample of VSS sources, employed by Tielens et al., but which also had arcsecond resolution, adequate for resolving practically all the sources in the sample (this was a significant improvement over the 4C sample of VSS sources observed by Tielens et al.\cite{1979A&AS...35..153T}, in which merely a quarter of those sources had in fact been resolved).\\

{\bf (b)  Spectral index - flux density [$\alpha$ (median) - S] relation for extragalactic sources selected at metre wavelengths}\\

By combining the BOOTY sample with other metre-wavelength samples having published quality measurements of both flux density and spectral index, and supplementing these data with their own Effelsberg measurements of a complete sample of 212 sources selected from the Molonglo sample defined at 408 MHz, it was shown by Gopal-Krishna \& Steppe\cite{1982A&A...113..150G} and Steppe \& Gopal-Krishna\cite{1984A&A...135...39S}, for the first time, that for metre-wavelength selected extragalactic radio sources, median spectral index peaks and attains its steepest value ($\alpha$ (median) = $-$ 0.90) around a flux density level of $\sim$ 1 Jy at 408 MHz, and it flattens to $-$ 0.8 at flux densities on either side of the peak, thus mimicking the well-known differential counts of extragalactic radio sources found in meter-wavelength surveys. Although the claimed statistically significant flattening of $\alpha$ (median) below S408 $\sim$ 1 Jy was contested in some subsequent studies (see, \cite{1986A&A...165...39K,1989AJ.....98..419V,1990A&AS...82...41K,2003A&A...404...57Z}), the original result has recently been validated in 3 independent investigations based on more sensitive and much larger sky surveys conducted at metre-wavelengths, that have since become available in the public domain\cite{2019RAA....19...96T,2018MNRAS.474.5008D,2023A&A...675L...3D}.

\begin{figure}[!t]
\centering{
\includegraphics[width=.9\columnwidth]{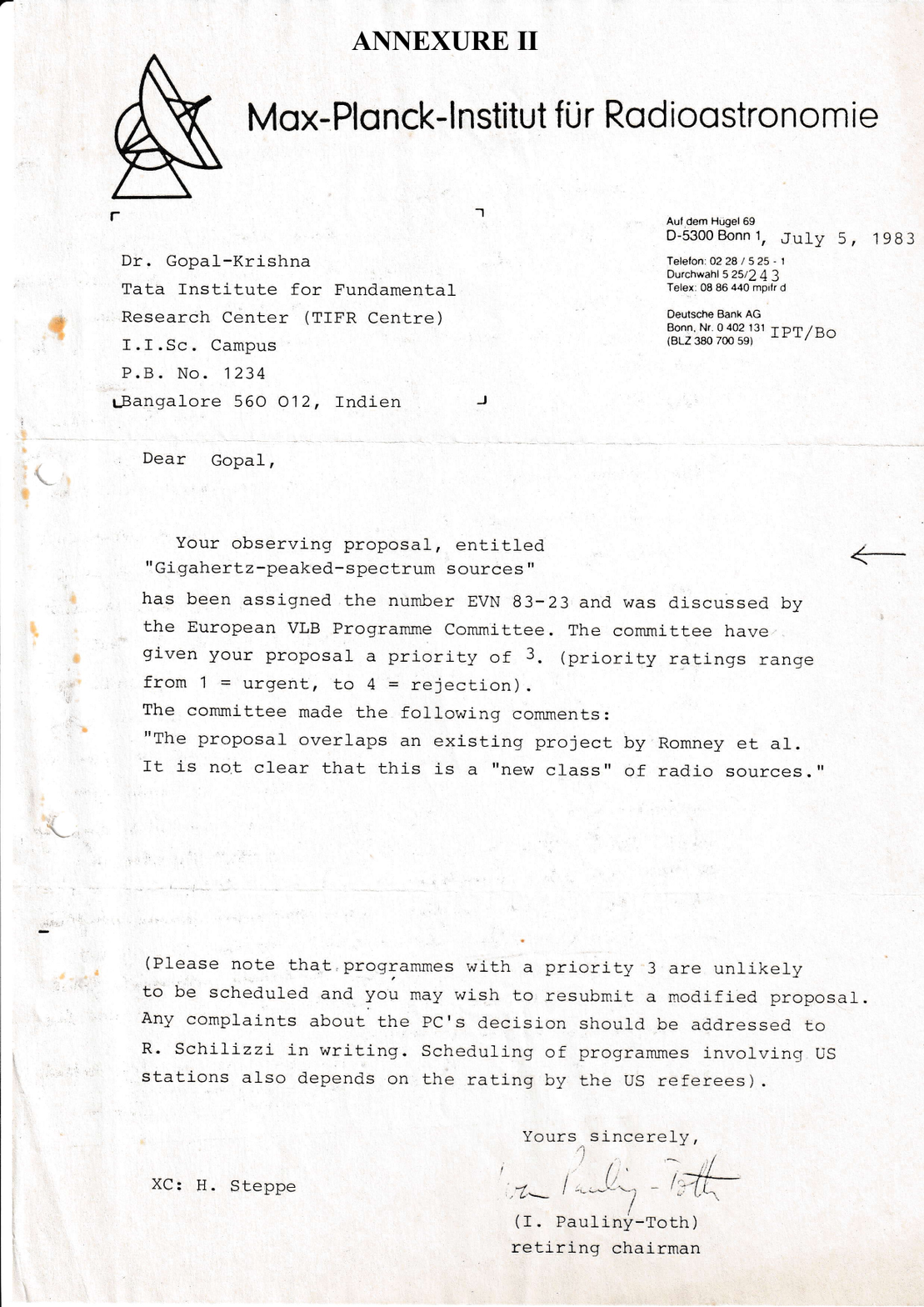}
}
\end{figure}
\begin{figure}[!t]
\centering{
\includegraphics[width=.9\columnwidth]{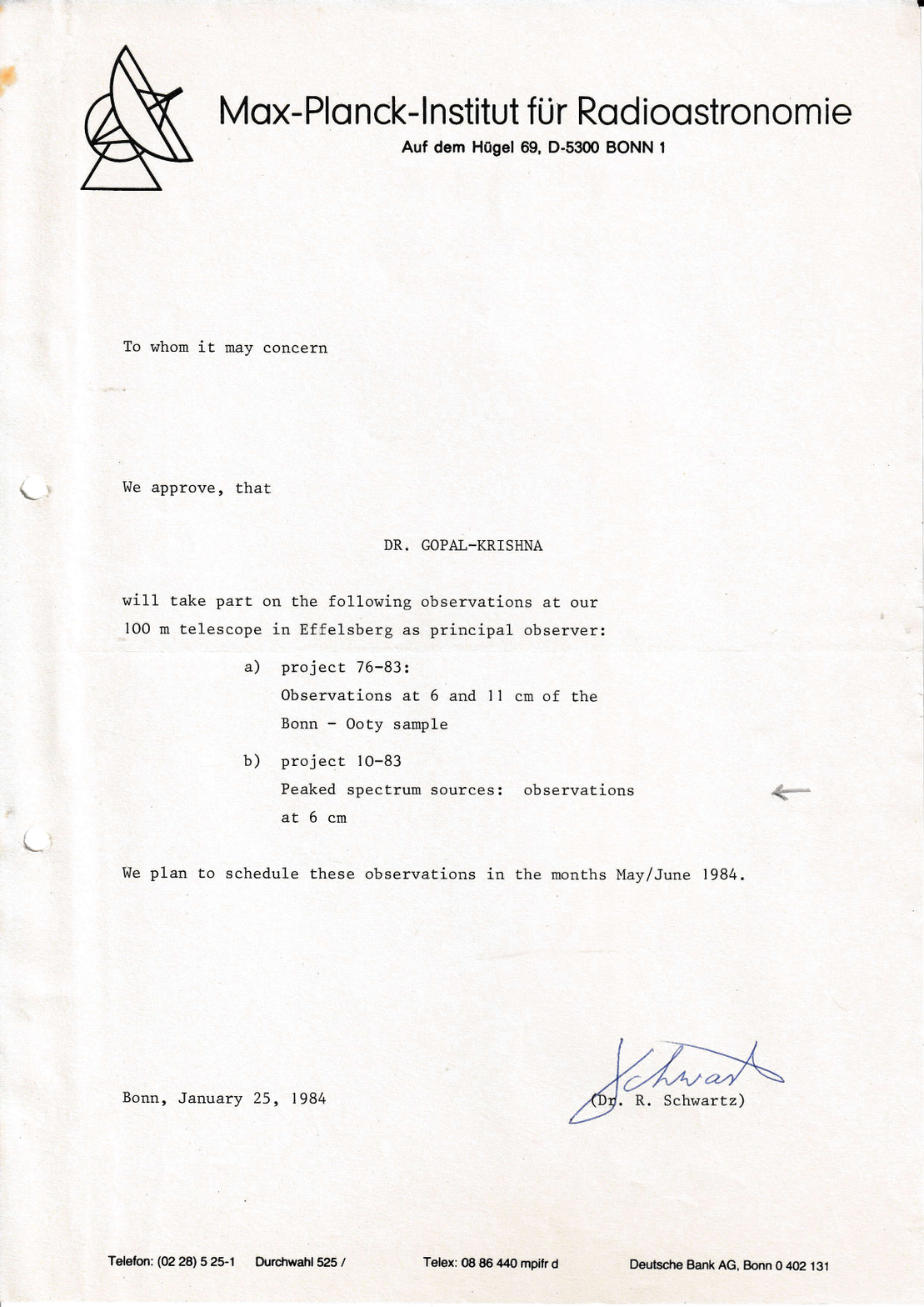}
}
\end{figure}
\begin{figure}[!t]
\centering{
\includegraphics[width=.9\columnwidth]{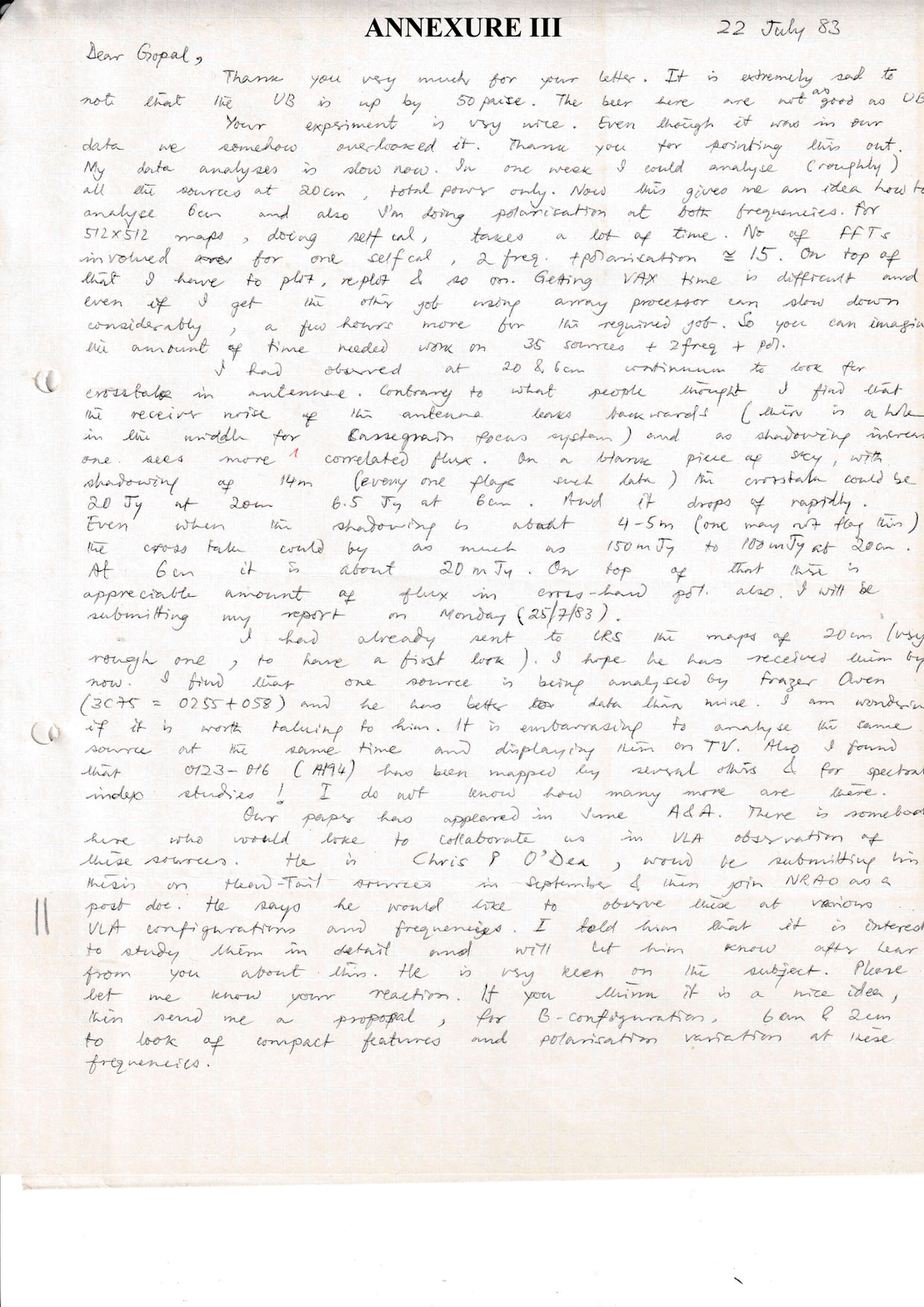}
}
\end{figure}
\begin{figure}[!t]
\centering{
\includegraphics[width=.9\columnwidth]{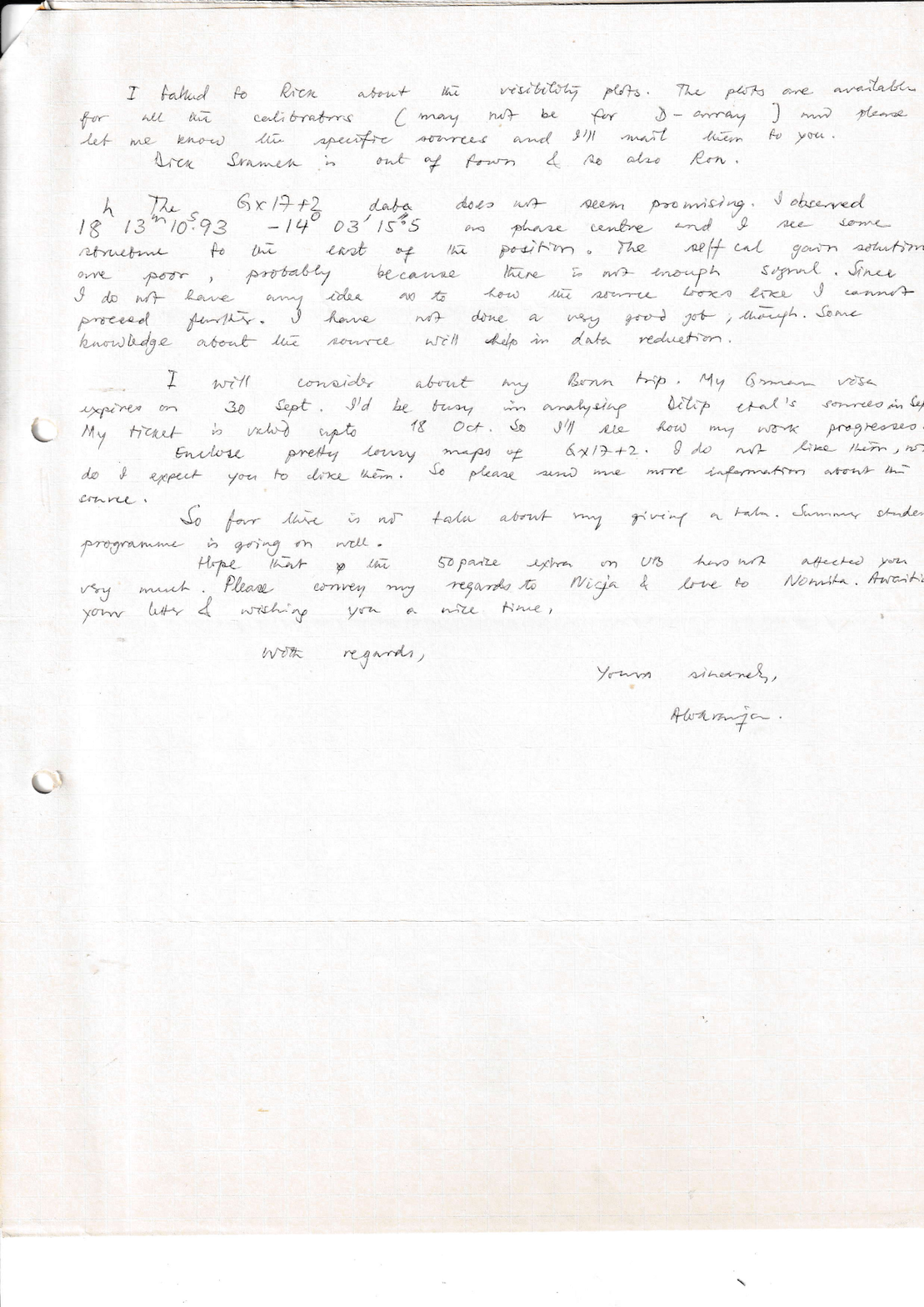}
}
\end{figure}



\section*{Acknowledgements}
The motivation behind this note was to shed light on and chronicle the induction of a major subclass of extragalactic radio sources which has come to be known as {\it gigahertz-peaked-spectrum} (GPS) sources.
The original intent, observations and coincidences that gave a kickstart to research activity in this now mature and thriving branch of extragalactic radio astronomy, are recounted and placed in the modern context, peering 4 decades down the memory lane, into an era when the research landscape lacked today's life-support-systems, such as e-mail and the various search engines that are now routinely available for easing the task of scouring the astronomy literature. 

I take immense pleasure in dedicating this article to the memory of my friend late Dr Hans Steppe of Kleinwalsertal (Austria). Collaboration with him at the Max-Planck-Institut f$\ddot{\mathrm{u}}$r Radioastronomie (Bonn) laid the foundation for a number of studies, some of which are documented in this article and the accompanying three Annexures, along with reminiscences of some remarkable coincidences that steered the course of events. I am also grateful to Profs. Ken Kellermann and Arno Witzel for the insightful support in course of the studies recounted in this article. The award of a Senior Scientist position by the Indian National Science Academy is also acknowledged. This article has benefitted from interaction with Dr. Krishan Chand.


\begin{thebibliography}{99} 
\bibitem{1969ApJ...157....1K}Kellermann, K. I., Pauliny-Toth, I. I. K. and Williams, P. J. S., The Spectra of Radio Sources in the Revised 3c Catalogue. The Astrophysical Journal, 1969, {\bf157}, 1.
\bibitem{1962MmRAS..68..163B} Bennett, A. S., The revised 3C catalogue of radio sources. Memoirs of the Royal Astronomical Society, 1962, {\bf68}, 163.
\bibitem{Pikelner1953DoSSR..88..229P}Pikel'ner, S., Akademia Nauk SSSR Doklady, 1953, {\bf88}, 229.
\bibitem{Shklovskii1953AZh....30...15S}Shklovskii, I. S., The Problem of Cosmic Radioemis-sion. $<$TITLE lang="ru$>$Problema KosmiRadioizluâ. Astronomicheskii Zhurnal, 1953, {\bf30}, 15-36.
 \bibitem{Shklovskii1955AZh....32..215S}Shklovskii, I. S., O Prirode Izluâ Radiogalaktiki NGC
44860 Prirode Izluâ Radiogalaktiki NGC 4486,On the
Nature of the Emission of Radiogalaxy NGC 4486. Astronomicheskii Zhurnal,
1955, {\bf32}, 215.
\bibitem{Kellermann1966AuJPh..19..195K} Kellermann, K. I., The radio source 1934-63. Australian Journal of Physics, 1966, {\bf19}, 195.
\bibitem{1963Natur.199..682B} Bolton, J. G., Gardner, F. F. and Mackey, M. B., A
Radio Source with a Very Unusual Spectrum. Nature, 1963, {\bf199}(4894), 682-683.
\bibitem{1965Natur.206..176S} Shklovsky, J., Possible Secular Variation of the Flux and Spectrum of Radio Emissions of Source 1934-63. Nature, 1965, {\bf206}(4980), 176-177.
\bibitem{2012arXiv1210.0984K} Kellermann, K. I., Early Parkes Observations of Planets and Cosmic Radio Sources. arXiv e-prints, 2012, arXiv:1210.0984.
\bibitem{2024arXiv:2406.13346}  Ballieux, F. J., Callingham, J. R., Röttgering, H. J. A. and Slob, M. M., Comparing extragalactic megahertz-peaked spectrum and gigahertz-peaked spectrum sources. arXiv e-prints, 2024, arXiv:2406.13346.
\bibitem{2000A&A...363..887D} Dallacasa, D., Stanghellini, C., Centonza, M. and Fanti, R., High frequency peakers. I. The bright sample. Astronomy and Astrophysics, 2000, {\bf363}, 887-900.
\bibitem{2015MNRAS.450.1477C} Coppejans, R., Cseh, D., Williams, W. L., van Velzen, S. and Falcke, H., Megahertz
peaked-spectrum sources in the Boötes field I - a route towards finding high-redshift AGN. Monthly Notices of the Royal Astronomical Society, 2015, {\bf450}(2), 1477-1485.
\bibitem{2021A&ARv..29....3O} O'Dea, C. P. and Saikia, D. J., Compact steep-spectrum and peaked-spectrum radio sources. Astronomy and Astrophysics Review, 2021, {\bf29}(1), 3.
\bibitem{Razin1960} Razin, V. A., Radiofizika, 1960, {\bf3}, 921.
\bibitem{1964SvA.....8..217K} Kardashev, N. S., Transmission of Information by Extraterrestrial Civilizations. Soviet Astronomy, 1964, {\bf8}, 217.
\bibitem{1972ApJ...173L..19P} Peterson, B. A. and Bolton, J. G., Redshifts of Southern Radio Sources. The Astrophysical Journal, 1972, \textbf{{173}}, L19.
\bibitem{1974ApJ...193..293R} Rogers, A. E. E., Hinteregger, H. F., Whitney, A. R., Counselman, C. C., Shapiro, I. I., Wittels, J. J. et al., The structure of radio sources 3C 273B and 3C 84 deduced from the ``closure'' phases and visibility amplitudes observed with three-element interferometers. The Astrophysical Journal, 1974, {\bf193}, 293-301.
\bibitem{1977Natur.269..764W} Wilkinson, P. N., Readhead, A. C. S., Purcell, G. H. and Anderson, B., Radio structure of 3C 147 determined by multi-element very long baseline interferometry. Nature, 1977, {\bf269}, 764-768.
\bibitem{1980ApJ...236...89P} Phillips, R. B. and Mutel, R. L., High-resolution observations of the compact radio sources CTD 93 and 3C 395 at 1671 megahertz. The Astrophysical Journal, 1980, {\bf236}, 89-98.
\bibitem{1982A&A...106...21P} Phillips, R. B. and Mutel, R. L., On symmetric structure in compact radio sources. Astronomy and Astrophysics, 1982, {\bf106},
21-24.
\bibitem{1953Natur.172..996J} Jennison, R. C. and Gupta, M. K. D., Fine Structure of the Extra-terrestrial Radio Source Cygnus I. Nature, 1953, {\bf172}(4387), 996-997.
\bibitem{1981ApJ...244...19P} Phillips, R. B. and Mutel, R. L., Milliaresecond structure of 0428+205, 1518+047, and 2050+364 at 1.67
GHz. The Astrophysical Journal, 1981, {\bf244}, 19-26.
\bibitem{1981AJ.....86.1600M} Mutel, R. L., Phillips, R. B. and Skuppin, R., The structure of DA 344 at 1.67 GHz. The Astronomical Journal, 1981, {\bf86}, 1600-1603.
\bibitem{1969ApJ...155L..71K} Kellermann, K. I. and Pauliny-Toth, I. I. K., The Spectra of Opaque Radio Sources. The Astrophysical Journal, 1969, {\bf155}, L71.
\bibitem{1992ApJ...396...62C} Conway, J. E., Pearson, T. J., Readhead, A. C. S., Unwin, S. C., Xu, W. and Mutel, R. L., The Compact Triples 0710+439 and 2352+495: A New Morphology of Radio Galaxy Nuclei. The Astrophysical Journal, 1992, {\bf396}, 62.
\bibitem{1994ApJ...432L..87W} Wilkinson, P. N., Polatidis, A. G., Readhead, A. C. S., Xu, W. and Pearson, T. J., Two-sided Ejection in Powerful Radio Sources: The Compact Symmetric Objects. The Astrophysical Journal, 1994, {\bf432}, L87.
\bibitem{2001A&A...369..380F} Fanti, C., Pozzi, F., Dallacasa, D., Fanti, R., Gregorini, L., Stanghellini, C. et al., Multi-frequency VLA observations of a new sample of CSS/GPS radio sources. Astronomy and Astrophysics, 2001, {\bf369}, 380-420.
\bibitem{Kiehlmann2024ApJ...961..241K} Kiehlmann, S., Readhead, A. C. S., O'Neill, S., Wilkinson, P. N., Lister, M. L., Liodakis, I. et al., Compact Symmetric Objects. II. Confirmation of a Distinct Population of High-luminosity Jetted Active Galaxies. The Astrophysical Journal, 2024, {\bf961}(2), 241.
\bibitem{1994cers.conf...17R} Readhead, A. C. S., Xu, W., Pearson, T. J., Wilkinson, P. N. and Polatidis, A. G., Compact Symmetric Objects. In Compact Extragalactic Radio Sources, edited by Zensus, J. A. and Kellermann, K. I., 1994, p. 17.
\bibitem{1995A&A...302..317F} Fanti, C., Fanti, R., Dallacasa, D., Schilizzi, R. T., Spencer, R. E. and Stanghellini, C., Are compact steep-spectrum sources young? Astronomy and Astrophysics, 1995, {\bf302}, 317.
\bibitem{1996ApJ...460..634R} Readhead, A. C. S., Taylor, G. B., Pearson, T. J. and Wilkinson, P. N., Compact Symmetric Objects and the Evolution of Powerful Extragalactic Radio Sources. The Astrophysical Journal, 1996, {\bf460}, 634.
\bibitem{2005ApJ...622..136G} Gugliucci, N. E., Taylor, G. B., Peck, A. B. and Giroletti, M., Dating COINS: Kinematic Ages for Compact Symmetric Objects. The Astrophysical Journal, 2005, {\bf622}(1), 136-148.
\bibitem{1997ApJ...487L.135R} Reynolds, C. S. and Begelman, M. C., Intermittant Radio Galaxies and Source Statistics. The Astrophysical Journal, 1997, {\bf487}(2), L135-L138.
\bibitem{2006A&A...450..945K} Kunert-Bajraszewska, M., Marecki, A. and Thomasson, P., FIRST-based survey of compact steep spectrum sources. IV. Multifrequency VLBA observations of very compact objects. Astronomy and Astrophysics, 2006, {\bf450}(3), 945-958.
\bibitem{1982ApJ...260L..27P} Peterson, B. A., Savage, A., Jauncey, D. L. and Wright, A. E., PKS 2000-330: a quasi-stellar radio source with a redshift of 3.78. The Astrophysical Journal, 1982, {\bf260}, L27-L29.
\bibitem{1981MNRAS.197..593D} Downes, A. J. B., Longair, M. S. and Perryman, M. A. C., High-resolution observations of faint radio
sources and the angular size-flux density relation. Monthly Notices of the Royal Astronomical Society, 1981, {\bf197}, 593-626.
\bibitem{1981A&AS...43..381K} Kapahi, V. K., Westerbork observations of radio sources in the 5GHz ``S4'' survey. Astronomy and Astrophysics Supplement Series, 1981, {\bf43}, 381-393.
\bibitem{1978MNRAS.185..123K} Kulkarni, V. K., Frequency dependence of radio-
source counts and spectral index distributions. Monthly Notices of the Royal Astronomical Society, 1978, {\bf185}, 123.
\bibitem{1971NPhS..230..185S} Swarup, G., Sarma, N. V. G., Joshi, M. N., Kapahi, V. K., Bagri, D. S., Damle, S. H. et al., Large Steerable Radio Telescope at Ootacamund, India. Nature Physical Science, 1971, {\bf230}(17), 185-188.
\bibitem{1971ApL.....9...53S} Swarup, G., Kapahi, V. K., Sarma, N. V. G., Gopal-Krishna, Joshi, M. N. and Rao, A. P., Lunar Occultation Observations of 25 Radio Sources Made with the
Ooty Radio Telescope: List 1. Astrophysical Letters, 1971, {\bf9}, 53.
\bibitem{1981A&AS...45..367K} K$\ddot{\mathrm{u}}$hr, H., Witzel, A., Pauliny-Toth, I. I. K. and Nauber, U., A Catalogue of Extragalactic Radio Sources Having Flux Densities Greater than 1-JY at 5-GHZ. Astronomy and Astrophysics Supplement Series, 1981, {\bf45}, 367.
\bibitem{Kuehr1981MPIfR} K$\ddot{\mathrm{u}}$hr, H., Nauber, U., Pauliny-Toth, I. I. K. and Witzel, A., Max-Planck-Institut f$\ddot{\mathrm{u}}$r Radioastronomie, 1981, preprint No. 55.
\bibitem{1983A&A...123..107G} Gopal-Krishna, Patnaik, A. R. and Steppe, H., A
sample of 25 extragalactic radio sources having a
spectrum peaked around 1 GHz. Astronomy and Astrophysics, 1983, {\bf123}, 107-110.
\bibitem{1985A&A...152...38S} Spoelstra, T. A. T., Patnaik, A. R. and Gopal-Krishna, A sample of 25 extragalactic radio sources having
a spectrum peaked around 1 GHz (list 2). Astronomy and Astrophysics, 1985, {\bf152}, 38-41.
\bibitem{1993A&A...271..101G} Gopal-Krishna and Spoelstra, T. A. T., A sample of gigahertz-peaked-spectrum radio sources: List 3. Astronomy and Astrophysics, 1993, {\bf271}, 101-103.
\bibitem{1982ApJ...255...39R} Rudnick, L. and Jones, T. W., Compact radio sources : the dependence of variability and polarization on
spectral shape. The Astrophysical Journal, 1982, {\bf255}, 39-47.
\bibitem{1984IAUS..110...15P} Pearson, T. J. and Readhead, A. C. S., VLBI survey of a complete sample of active nuclei and quasars. In VLBI and Compact Radio Sources, edited by Fanti, R., Kellermann, K. I.
and Setti, G., 1984, vol. {\bf110}, pp. 15–24.
\bibitem{1998PASP..110..493O} O'Dea, C. P., The Compact Steep-Spectrum and Gigahertz Peaked-Spectrum Radio Sources. Publications of the Astronomical Society of the Pacific, 1998, {\bf110}(747), 493-532.
\bibitem{1998A&AS..131..303S} Stanghellini, C., O'Dea, C. P., Dallacasa, D., Baum, S. A., Fanti, R. and Fanti, C., A complete sample
of GHz-peaked-spectrum radio sources and its radio
properties. Astronomy and Astrophysics Supplement Series, 1998, {\bf131}, 303-315.
\bibitem{2000MNRAS.319..445S} Snellen, I. A. G., Schilizzi, R. T., Miley, G. K., de Bruyn, A. G., Bremer, M. N. and Röttgering, H. J. A., On the evolution of young radio-loud AGN. Monthly Notices of the Royal Astronomical Society,
2000, {\bf319}(2), 445-456.
\bibitem{2016MNRAS.459.2455C} Coppejans, R., Cseh, D., van Velzen, S., Falcke, H., Intema, H. T., Paragi, Z. et al., What are the megahertz peaked-spectrum sources? Monthly Notices of the Royal Astronomical Society, 2016, {\bf459}(3), 2455-2471.
\bibitem{1998A&A...329..845C} Carvalho, J. C., The evolution of GHz-peaked-spectrum radio sources. Astronomy and Astrophysics, 1998, {\bf329}, 845-852.
\bibitem{Kunert2010MNRAS.408.2279K} Kunert-Bajraszewska, M. and Labiano, A., A survey of low-luminosity compact sources and its implication for the evolution of radio-loud active galactic nuclei - II. Optical analysis. Monthly Notices of the Royal Astronomical Society, 2010, {\bf408}, 2279-2289.
\bibitem{2012ApJ...760...77A} An, T. and Baan, W. A., The Dynamic Evolution of Young Extragalactic Radio Sources. The Astrophysical Journal, 2012, {\bf760}(1), 77.
\bibitem{1998A&A...336L..37O} Owsianik, L., Conway, J. E. and Polatidis, A. G., Renewed Radio Activity of Age 370 years in the Extragalactic Source 0108+388. Astronomy and Astrophysics, 1998, {\bf336}, L37-L40.
\bibitem{1998A&A...337...69O} Owsianik, I. and Conway, J. E., First detection of hotspot advance in a Compact Symmetric Object. Evidence for a class of very young extragalactic radio sources. Astronomy and Astrophysics, 1998, {\bf337}, 69-79.
\bibitem{2021AN....342.1151O} Orienti, M. and Dallacasa, D., Young radio sources: From newly born to short-lived objects. Astronomische Nachrichten, 2021, {\bf342}(1151), 1151-1154.
\bibitem{1995PNAS...9211399G} Gopal-Krishna, The Case for Unification. Proceedings of the National Academy of Science, 1995, {\bf92}(25), 11399-11406.
\bibitem{2009ApJ...697..1621G} Gergely, L. A. and Biermann, P. L., The spin-flip phenomenon in supermassive black hole binary mergers. The Astrophysical Journal, 2009, {\bf697}, 1621-1633.
\bibitem{1982ApJ...262..529H} Heckman, T. M., Miley, G. K., Balick, B., van Breugel, W. J. M. and Butcher, H. R., An optical and radio investigation of the radio galaxy 3C 305. The Astrophysical Journal, 1982, {\bf262}, 529-553.
\bibitem{1984AJ.....89....5V} van Breugel, W., Miley, G. and Heckman, T., Studies of kiloparsec-scale, steep-spectrum radio cores. I. VLA maps. The Astronomical Journal, 1984, {\bf89}, 5-22.
\bibitem{2022ApJ...927L..24G} Gupta, N., Srianand, R., Momjian, E., Shukla, G., Combes, F., Krogager, J. K. et al., HI Gas Playing Hide-and-seek around a Powerful FRI-type Quasar at z 2.1. The Astrophysical Journal, 2022, {\bf927}(2), L24.
\bibitem{1991ApJ...373..325G} Gopal-Krishna and Wiita, P. J., Gaseous Halos of Elliptical Galaxies, the Cosmic Evolution of Their Radio Sizes, and the Phenomenon of Compact Steep-Spectrum Sources. The Astrophysical Journal, 1991, {\bf373}, 325.
\bibitem{1982MNRAS.198..843P} Peacock, J. A. and Wall, J. V., Bright extragalactic radio sources at 2.7 GHz- II. Observations with the Cambridge 5-km telescope. Monthly Notices of the Royal Astronomical Society, 1982, {\bf198}, 843-860.
\bibitem{2023Galax..11...24M} Morganti, R., Murthy, S., Guillard, P., Oosterloo, T. and Garcia-Burillo, S., Young Radio Sources Expanding in Gas-Rich ISM: Using Cold Molecular Gas to Trace Their Impact. Galaxies, 2023, {\bf11}(1), 24.
\bibitem{Kukreti2024arXiv240706265K} Kukreti, P. and Morganti, R., Connecting the radio AGN life cycle to feedback: Ionised gas is more disturbed in young radio AGN. arXiv e-prints, 2024, arXiv:2407.06265.
\bibitem{1990A&A...231..333F} Fanti, R., Fanti, C., Schilizzi, R. T., Spencer, R. E., Rendong, N., Parma, P. et al., On the nature of compact steep spectrum radio sources. Astronomy and Astrophysics, 1990, {\bf231}, 333-346.
\bibitem{1997AJ....113..148O} O'Dea, C. P. and Baum, S. A., Constraints on Radio Source Evolution from the Compact Steep Spectrum and GHz Peaked Spectrum Radio Sources. The Astronomical Journal, 1997, {\bf113}, 148-161.
\bibitem{2009AN....330..214D} de Vries, N., Snellen, I. A. G., Schilizzi, R. T. and Mack, K. H., Further evidence for synchrotron self-absorption from the CORALZ sample of young radio-loud AGN. Astronomische Nachrichten, 2009, {\bf330}(2), 214.
\bibitem{Jeyakumar2016MNRAS.458.3786J} Jeyakumar, S., Numerical calculations of spectral turnover and synchrotron self-absorption in CSS and GPS radio sources. Monthly Notices of the Royal Astronomical Society, 2016, {\bf458}(4), 3786-3794.
\bibitem{1996ApJ...460..612R} Readhead, A. C. S., Taylor, G. B., Xu, W., Pearson, T. J., Wilkinson, P. N. and Polatidis, A. G., The Statistics and Ages of Compact Symmetric Objects. The Astrophysical Journal, 1996, {\bf460}, 612.
\bibitem{Tingay2003AJ....126..723T}  Tingay, S. J. and de Kool, An Investigation of Synchrotron Self-absorption and Free-Free Absorption Models in Explanation of the Gigahertz-peaked Spectrum of PKS 1718-649. The Astronomical Journal, 2003, {\bf126}(2), 723-733.
\bibitem{1997ApJ...485..112B} Bicknell, G. V., Dopita, M. A. and O'Dea, C. P. O., Unification of the Radio and Optical Properties of Gigahertz Peak Spectrum and Compact Steep-Spectrum Radio Sources. The Astrophysical Journal, 1997, {\bf485}(1), 112–124.
\bibitem{1999ApJ...521..103P} Peck, A. B., Taylor, G. B. and Conway, J. E., Obscuration of the Parsec-Scale Jets in the Compact Symmetric Object 1946+708. The Astrophysical Journal, 1999, {\bf521}(1), 103-111.
\bibitem{2003ApJ...584..135L} Lister, M. L., Kellermann, K. I., Vermeulen, R. C., Cohen, M. H., Zensus, J. A. and Ros, E., 4C +12.50: A Superluminal Precessing Jet in the Recent Merger System IRAS 13451+1232. The Astrophysical Journal, 2003, {\bf584}(1), 135-146.
\bibitem{2014ApJ...780..178M} Marr, J. M., Perry, T. M., Read, J., Taylor, G. B. and Morris, A. O., Multi-frequency Optical-depth Maps and the Case for Free-Free Absorption in Two Compact Symmetric Radio Sources: The CSO Candidate J1324 + 4048 and the CSO J0029 + 3457. The Astrophysical Journal, 2014, {\bf780}(2), 178.
\bibitem{2015ApJ...809..168C} Callingham, J. R., Gaensler, B. M., Ekers, R. D., Tin-gay, S. J., Wayth, R. B., Morgan, J. et al., Broadband Spectral Modeling of the Extreme Gigahertz-peaked Spectrum Radio Source PKS B0008-421. The Astrophysical Journal, 2015, {\bf809}(2), 168.
\bibitem{2018MNRAS.475.3493B} Bicknell, G. V., Mukherjee, D., Wagner, A. Y., Suther- land, R. S. and Nesvadba, N. P. H., Relativistic jet feedback - II. Relationship to gigahertz peak spectrum and compact steep spectrum radio galaxies. Monthly Notices of the Royal Astronomical Society, 2018, {\bf475}(3), 3493-3501.
\bibitem{Sobolewska2019ApJ...871...71S} Sobolewska, M., Siemiginowska, A., Guainazzi, M., Hardcastle, M., Migliori, G., Ostorero, L. and Stawarz, Ł., The Impact of the Environment on the Early Stages of Radio Source Evolution. The Astrophysical Journal, 2019, {\bf871}, 71.
\bibitem{2019MNRAS.489.3506M} Mhaskey, M., Gopal-Krishna, Paul, S., Dabhade, P., Salunkhe, S., Bhagat, S. et al., GMRT observations of a first sample of `Extremely Inverted Spectrum Extragalactic Radio Sources (EISERS)' candidates in the Northern sky. Monthly Notices of the Royal Astronomical Society, 2019, {\bf489}(3), 3506-3518.
\bibitem{Shao2022A&A...659A.159S} Shao, Y., Wagg, J., Wang, R., Momjian, E., Carilli, Chris L., Walter, F., Riechers, Dominik A., Intema, Huib T., Weiss, A., Brunthaler, A. and Menten, Karl M., The radio spectral turnover of radio-loud quasars at z $>$ 5. Astronomy \& Astrophysics, 2022, {\bf659}, A159.
\bibitem{2019A&A...628A..56K} Keim, M. A., Callingham, J. R. and Röttgering, H. J. A., Extragalactic megahertz-peaked spectrum radio sources at milliarcsecond scales. Astronomy and Astrophysics, 2019, {\bf628}, A56.
\bibitem{1979ApJ...228...27M} Marscher, A. P., Absorption models for low-frequency variability in compact radio sources. The Astrophysical Journal, 1979, {\bf228}, 27-33.
\bibitem{1980PASA....4...70M} McAdam, W. B., Variable Sources at 408-MHZ. Publications of the Astronomical Society of Australia, 1980, {\bf4}, 70.
\bibitem{1981ApJ...247..419U} Ulvestad, J. S., Wilson, A. S. and Sramek, R. A., Radio structures of Seyfert galaxies. II. The Astrophysical Journal, 1981, {\bf247}, 419-442.
\bibitem{1984ApJ...277...82V} van Breugel, W., Heckman, T., Butcher, H. and Miley, G., Extended optical line emission from 3C 293: radio jets propagating through a rotating gaseous disk. The Astrophysical Journal, 1984, {\bf277}, 82-91.
\bibitem{1963Natur.199..682S} Slish, V. I., Angular Size of Radio Stars. Nature, 1963, {\bf199}(4894), 682.
\bibitem{1979rpa..book.....R} Rybicki, G. B. and Lightman, A. P., Radiative processes in astrophysics, 1979.
\bibitem{1963Natur.200...56W} Williams, P. J. S., Absorption in Radio Sources of High Brightness Temperature. Nature, 1963, {\bf200}(4901), 56-57.
\bibitem{2014MNRAS.443.2824G} Gopal-Krishna, Sirothia, S. K., Mhaskey, M., Ranadive, P., Wiita, P. J., Goyal, A. et al., Extragalactic radio sources with sharply inverted spectrum at metre wavelengths. Monthly Notices of the Royal Astronomical Society, 2014, {\bf443}(3), 2824-2829. 
\bibitem{2019MNRAS.485.2447M} Mhaskey, M., Gopal-Krishna, Dabhade, P., Paul, S., Salunkhe, S. and Sirothia, S. K., GMRT observations of extragalactic radio sources with steeply inverted spectra. Monthly Notices of the Royal Astronomical Society, 2019, {\bf485}(2), 2447-2456.
\bibitem{2021A&A...648A.104D} de Gasperin, F., Williams, W. L., Best, P., Brüggen, M., Brunetti, G., Cuciti, V. et al., The LOFAR LBA Sky Survey. I. Survey description and preliminary data release. Astronomy and Astrophysics, 2021, {\bf648}, A104.
\bibitem{1987A&AS...69...91S} Singal, A. K., Ooty lunar occultation survey of radio sources. Astronomy and Astrophysics Supplement Series, 1987, {\bf69}(1), 91-115.
\bibitem{1965MmRAS..69..183P} Pilkington, J. D. H. and Scott, J. F., A survey of radio sources between declinations 20° and 40°. Memoirs of the Royal Astronomical Society, 1965, {\bf69}, 183.
\bibitem{1967MmRAS..71...49G} Gower, J. F. R., Scott, P. F. and Wills, D., A survey of radio sources in the declination ranges-07° to 20° and 40° to 80°. Memoirs of the Royal Astronomical Society, 1967, {\bf71}, 49.
\bibitem{1973AJ.....78.1023J} Joshi, M. N., Kapahi, V. K., Gopal-Krishna, Sarnia, N. V. G. and Swarup, G., Occultations of 50 radio sources at 327 MHz. The Astronomical Journal, 1973, {\bf78}, 1023.
\bibitem{1973AJ.....78..673K} Kapahi, V. K., Joshi, M. N., Subrahmanya, C. R. and Gopal-Krishna, Lunar-occultation studies of 4C radio sources at 327 MHz. The Astronomical Journal, 1973, {\bf78}, 673.
\bibitem{1974AJ.....79..515K} Kapahi, V. K., Joshi, M. N. and Sarma, N. V. G., Ooty Occultations of 76 radio sources. The Astronomical Journal, 1974, {\bf79}, 515.
\bibitem{1979MmASI...1....2S} Subrahmanya, C. R. and Gopal-Krishna, Ooty Occultations of 100 Radio Sources at 327-MHZ - List Six. Memoirs of the Astronomical Society of India, 1979, {\bf1}, 2.
\bibitem{1980MmASI...1...49J} Joshi, M. N. and Singal, A. K., Ooty Lunar Occultation Survey - List 9. Memoirs of the Astronomical Society of India, 1980, {\bf1}, 49.
\bibitem{1979A&AS...35..153T} Tielens, A. G. G. M., Miley, G. K. and Willis, A. G., Westerbork Observations of 4C Sources with Steep Radio Spectra. Astronomy and Astrophysics Supplement Series, 1979, {\bf35}, 153.
\bibitem{2008A&ARv..15...67M} Miley, G. and Breuck, C. D., Distant radio galaxies and their environments. Astronomy and Astrophysics Review, 2008, {\bf15}(2), 67-144. 
\bibitem{1981A&A...101..315G} Gopal-Krishna and Steppe, H., Extragalactic radio sources with very steep decimetre wave spectrum. Astronomy and Astrophysics, 1981, {\bf101}(3), 315-319.
\bibitem{1982A&A...113..150G} Gopal-Krishna and Steppe, H., Spectral index - Flux density relation for extragalactic radio sources found in metre-wavelength surveys. Astronomy and Astrophysics, 1982, {\bf113}, 150-154.
\bibitem{1984A&A...135...39S} Steppe, H. and Gopal-Krishna, The spectral index/flux density relation for extragalactic radio sources found in metre-wavelength surveys: an updating. Astronomy and Astrophysics, 1984, {\bf135}, 39-44.
\bibitem{1986A&A...165...39K} Kapahi, V. K. and Kulkarni, V. K., The spectral index-flux density relation at 408 MHz and the cosmological evolution of extragalactic radio sources. Astronomy and Astrophysics, 1986, {\bf165}, 39-44.
\bibitem{1989AJ.....98..419V} Vigotti, M., Grueff, G., Perley, R., Clark, B. G. and Bridle, A. H., Structures, Spectral Indexes, and Optical Identifications of Radio Sources Selected from the B3 Catalogue. The Astronomical Journal, 1989, {\bf98}, 419.
\bibitem{1990A&AS...82...41K} Kulkarni, V. K., Mantovani, F. and Pauliny-Toth, I. I. K., Spectral index distribution for a sample of sources from the B3 catalogue. Astronomy and Astrophysics Supplement Series, 1990, {\bf82}, 41-55.
\bibitem{2003A&A...404...57Z} Zhang, X., Reich, W., Reich, P. and Wielebinski, R., New results on the spectral index-flux density relation from the WENSS/NVSS catalogs. Astronomy and Astrophysics, 2003, {\bf404}, 57-62.
\bibitem{2019RAA....19...96T} Tiwari, P., Radio spectral index from NVSS and TGSS. Research in Astronomy and Astrophysics, 2019, {\bf19}(7), 096.
\bibitem{2018MNRAS.474.5008D} de Gasperin, F., Intema, H. T. and Frail, D. A., A radio spectral index map and catalogue at 147-1400 MHz covering 80 per cent of the sky. Monthly Notices of the Royal Astronomical Society, 2018, {\bf474}(4), 5008-5022.
\bibitem{2023A&A...675L...3D} Dabhade, P. and Gopal-Krishna, The spectral index-flux density relation for extragalactic radio sources selected at metre and decametre wavelengths. Astronomy and Astrophysics, 2023, {\bf675}, L3.




\end{thebibliography}



\end{document}